\documentclass[twocolumn,showpacs,amsfonts,aps,prc,floatfix,%
               superscriptaddress]{revtex4} 

\usepackage{amsmath}
\usepackage{bm}
\usepackage{graphicx}
\usepackage{color}

\newcommand{\be}[1]{\begin{equation}\label{#1}}
\newcommand{\ee}{\end{equation}}

\newcommand{\eq}{{\,=\,}}

\begin{document}

\title{Systematic parameter study of hadron spectra and elliptic flow 
from viscous hydrodynamic simulations of Au+Au collisions at 
$\sqrt{s_\mathrm{NN}} = 200$\,GeV}
\date{\today}

\author{Chun Shen}
\email[Correspond to\ ]{shen@mps.ohio-state.edu}
\affiliation{Department of Physics, The Ohio State University, 
  Columbus, OH 43210-1117, USA}
\author{Ulrich Heinz}
\email[Email:\ ]{heinz@mps.ohio-state.edu}
\affiliation{Department of Physics, The Ohio State University, 
  Columbus, OH 43210-1117, USA}
\author{Pasi Huovinen}
\email[Email:\ ]{huovinen@th.physik.uni-frankfurt.de}
\affiliation{Institut f\"ur Theoretische Physik, Johann Wolfgang 
Goethe-Universit\"at, Max-von-Laue-Stra\ss e 1, 60438 Frankfurt 
am Main, Germany}
\author{Huichao Song}
\email[Email:\ ]{HSong@LBL.gov}
\affiliation{Department of Physics, The Ohio State University, 
  Columbus, OH 43210-1117, USA}
\affiliation{Lawrence Berkeley National Laboratory, 1 Cyclotron Road,
MS70R0319, Berkeley, CA 94720, USA}

\begin{abstract}
Using the (2+1)-dimensional viscous hydrodynamic code {\tt VISH2+1}
\cite{Song:2007fn,Song:2007ux,Song:2009gc}, we present systematic studies
of the dependence of pion and proton transverse momentum spectra and
their elliptic flow in $200\,A$\,GeV Au+Au collisions on the parameters 
of the hydrodynamic model (thermalization time, initial entropy density 
distribution, decoupling temperature, equation of state, and specific 
shear viscosity $\eta/s$). We identify a tension between the slope of the
proton spectra, which (within hydrodynamic simulations that assume a 
constant shear viscosity to entropy density ratio) prefer larger $\eta/s$
values, and the slope of the $p_T$-dependence of charged hadron elliptic
flow, which prefers smaller values of $\eta/s$. Changing other model
parameters does not appear to permit dissolution of this tension.
\end{abstract}


\maketitle

\section{Introduction}
\label{sec1}

After experiments at the Relativistic Heavy Ion Collider (RHIC) 
\cite{Arsene:2004fa,Back:2004je,Adams:2005dq,Adcox:2004mh} and their
theoretical analysis \cite{Heinz:2001xi,Kolb:2003dz,Gyulassy:2004zy}
established that the quark-gluon plasma (QGP) created in ultra-relativistic
heavy-ion collisions is strongly coupled and behaves like an almost
ideal fluid (``perfect liquid'') with very small viscosity, interest in
the theoretical and phenomenological determination of the QGP transport
parameters, in particular its specific shear viscosity $\eta/s$ (i.e.
the ratio between its shear viscosity $\eta$ and entropy density $s$),
soared (see \cite{Romatschke:2009im,Heinz:2009xj} for recent reviews).
In principle, it should be possible to extract this quantity from heavy-ion
collision experiments, by comparing the measured hadron spectra and 
their azimuthal anisotropies (in particular their elliptic flow)
with theoretical simulations of the collision dynamics which treat
the QGP shear viscosity as an adjustable parameter 
\cite{Luzum:2008cw,Lacey:2010fe}. In practice, this is a complex and
difficult task that requires careful and highly constrained simulations 
of all dynamical stages of the collision that sandwich the viscous 
hydrodynamic expansion of the QGP between non-equilibrium phases 
describing (i) the initial geometry and early evolution of the fireball 
before its thermalization and (ii) the final kinetic hadron rescattering 
stage after its hadronization \cite{Song:2008hj,Song:2009gc}.

The present work is a contribution to help prepare the path for such
a phenomenological extraction of $(\eta/s)_\mathrm{QGP}$. It employs
viscous hydrodynamics to describe the fireball evolution, sidestepping
the issues related to early and late non-equilibrium evolution by
replacing the output from the (hypothetical) early non-equilibrium 
evolution model by initial conditions for the hydrodynamic stage (to
be adjusted {\em post facto} to final hadron spectra and multiplicities in
central collisions \cite{Kolb:2003dz}), and the late-stage hadronic 
rescattering and kinetic freeze-out by a sudden transition from viscous 
fluid to free-streaming particles, using the Cooper-Frye algorithm 
\cite{Cooper:1974mv} along a hypersurface of constant temperature 
$T_\mathrm{dec}$. This generalizes analogous attempts to describe 
experimental data from $200\,A$\,GeV Au+Au collisions at RHIC with
ideal fluid dynamics \cite{Hirano:2002ds,Kolb:2002ve,Kolb:2003dz,%
Huovinen:2005gy,Huovinen:2006jp,Huovinen:2007xh,Eskola:2007zc,%
Broniowski:2008vp} to the case of viscous fluid dynamics. Related
work has already been reported in \cite{Luzum:2008cw,Bozek:2009ty};
what distinguishes the present study from these earlier papers is
that we use a state-of-the-art equation of state that matches the 
latest Lattice QCD data \cite{Cheng:2007jq,Bazavov:2009zn} at high 
temperatures to a realistic, chemically non-equilibrated hadron resonance 
gas at low temperatures. The construction of this EoS is described in 
\cite{Huovinen:2009yb}, except that we here implement chemical freeze-out 
of the stable hadron yield ratios at $T_\mathrm{chem}\eq165$\,MeV, by 
imposing appropriate temperature dependent non-equilibrium chemical 
potentials for each hadron species below $T_\mathrm{chem}$ 
\cite{Bebie:1991ij,Hirano:2002ds,Kolb:2002ve,Teaney:2002aj,Huovinen:2007xh}. 
This ensures that the final hadron yield ratios from our simulations are 
consistent with their measured values which indicate chemical equilibrium 
at temperature $T_\mathrm{chem}{\,\approx\,}160-170$\,MeV
\cite{Adams:2005dq,BraunMunzinger:2001ip,Andronic:2008gu}.

The purpose of this study is {\it not} a detailed viscous hydrodynamic 
fit to the RHIC data; its goal is rather to build intuition for systematic
trends and parameter dependences that will be useful in forthcoming more 
ambitious fit attempts. One feature that disqualifies the present model 
study from being taken too seriously in comparison with the experimental 
data is our assumption of a constant (i.e. temperature independent) 
specific entropy $\eta/s$. While $\eta/s$ is probably small in the QGP 
phase \cite{Luzum:2008cw,Lacey:2010fe,Lacey:2006pn}, possibly close to 
the KSS bound $\left(\frac{\eta}{s}\right)_\mathrm{KSS}\eq\frac{1}{4\pi}$ 
\cite{Policastro:2001yc,Kovtun:2004de}, it is expected to increase 
dramatically in the late dilute hadronic phase 
\cite{Csernai:2006zz,Demir:2008tr}. This can have important consequences 
for the evolution of flow in relativistic heavy-ion collisions 
\cite{Bozek:2009dw} which will be studied in a separate paper 
\cite{ChunShen2010}.

\section{Hydrodynamic equations, initial and final conditions} 
\label{sec2}

In this work, we use viscous hydrodynamics to simulate the 
collision system by solving the second-order Israel-Stewart 
equations as described in Ref.~\cite{Song:2008si}. The energy-momentum
tensor of the fluid is decomposed as 
\begin{equation}
T^{\mu \nu} = e u^\mu u^\nu - (p + \Pi) \Delta^{\mu \nu} + \pi^{\mu \nu}
\label{eq1}
\end{equation}
where $e$ is the local energy density, $p$ is the thermal equilibrium
pressure (given by the equation of state $p(e)$, see below), $u^\mu$ is 
the local flow 4-velocity, $\Delta^{\mu\nu}\eq{g}^{\mu\nu}{-}u^\mu
u^\nu$ is the spatial projector in the local fluid rest frame, $\Pi$ is 
the bulk viscous pressure (which we set to zero in this paper, assuming 
that effects from bulk viscosity can be ignored relative to those caused 
by shear viscosity \cite{Song:2009rh}), and $\pi^{\mu \nu}$ is the traceless 
and symmetric shear pressure tensor satisfying $u_\mu\pi^{\mu \nu}\eq0$. The 
equations of motion are the hydrodynamic equations
\begin{equation}
d_\mu T^{\mu \nu} = 0, 
\label{ISeq1}
\end{equation}
where $d_\mu$ denotes the covariant derivative in curvilinear 
$(\tau,x,y,\eta)$ coordinates (see \cite{Heinz:2005bw,Song:2008si} for 
details), coupled to the Israel-Stewart \cite{Israel:1979wp,Muronga:2001zk,%
Muronga:2003ta,Song:2008si} evolution equations for the viscous pressure 
components:
\begin{eqnarray}
\Delta^{\mu\alpha} \Delta^{\nu\beta} \dot{\pi}_{\alpha\beta} =  
- \frac{\pi^{\mu\nu}{-}2 \eta \sigma^{\mu\nu}}{\tau_{\pi}}  
- \frac{\pi^{\mu\nu}}{2}\frac{\eta T}{\tau_\pi} d_\lambda 
\bigg( \frac{\tau_\pi}{\eta T}u^\lambda \bigg).
\label{ISeq2}
\end{eqnarray}
The dot on the left hand side stands for the local comoving time derivative
$D\eq{u^\mu d_\mu}$, $\eta$ is the shear viscosity, 
$\sigma^{\mu\nu}\eq\nabla^{\left\langle\mu\right.}u^{\left.\nu\right\rangle}$
is the velocity shear tensor (see \cite{Heinz:2005bw,Song:2008si} for
notation), and $\tau_\pi$ is the microscopic relaxation time that controls 
the evolution of $\pi^{\mu\nu}$ (we take $\tau_\pi\eq3\frac{\eta}{sT}$ 
\cite{Song:2008si}).

The equations are solved numerically in the two transverse spatial directions 
and time, using the (2+1)-dimensional hydrodynamic code {\tt VISH2+1} 
\cite{Song:2007fn,Song:2007ux,Song:2009gc}, assuming boost-invariant 
longitudinal expansion along the beam direction. The net baryon density and 
heat conductivity are set to zero.

To initialize the hydrodynamic evolution we must specify the starting 
time $\tau_0$ at which the system is sufficiently close to local thermal
equilibrium for viscous hydrodynamics to be applicable, initial
energy density and velocity profiles, and the initial viscous
pressure tensor $\pi^{\mu\nu}$. We here consider $\tau_0$ as a tunable 
parameter and vary it between 0.2 and 0.8\,fm/$c$ in order to study 
how it affects the final hadron spectra and elliptic flow.

\begin{figure}[h]
 \includegraphics[width=\linewidth,clip=]{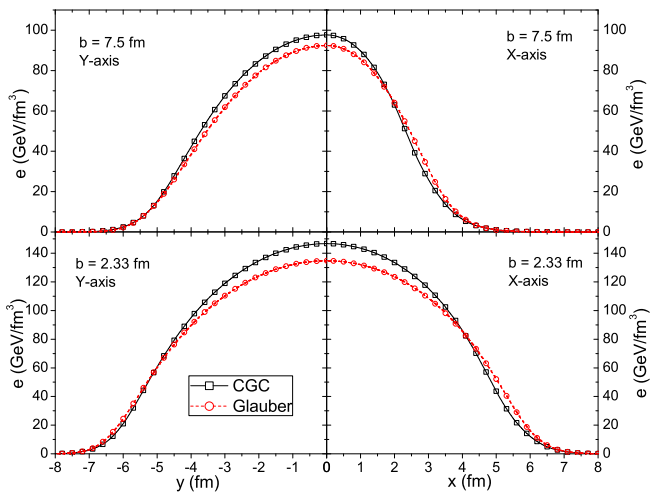}
 \caption{(Color online) A comparison of initial energy density profiles 
 at $\tau_0 = 0.4$\,fm/$c$ for ``central'' ($b\eq2.33$\,fm, bottom) and 
 ``peripheral'' ($b\eq7.5$\,fm, top) Au+Au collisions from the Glauber 
 and CGC-fKLN models. Shown are cuts along the $x$-axis (right panels) 
 and $y$-axis (left panels). The two profiles are normalized to the 
 same total entropy at $b\eq2.33$\,fm, using the EoS s95p-PCE to convert 
 energy to entropy density.}
 \label{F1}
\end{figure}

For the initial energy density profile we study both Glauber 
\cite{Kolb:1999it,Kolb:2000sd,Kolb:2001qz,Song:2007ux} and color glass 
condensate (CGC-fKLN) initializations \cite{Kharzeev:2000ph,Kharzeev:2002ei,%
Drescher:2006pi,Drescher:2006ca}, in the optical limit (i.e. without
accounting for event-by-event fluctuations \cite{Hirano:2009ah,%
Holopainen:2010qz,Schenke:2010rr,Petersen:2010md,Andrade:2008xh}).
Figure \ref{F1} shows a comparison of typical initial energy density 
profiles generated from Glauber and CGC initializations. In the Glauber 
model we assume a mixture of 85\% wounded nucleon and 15\% binary collision 
contributions to the entropy production \cite{Hirano:2005xf}. For the CGC 
model we assume that the entropy density is proportional to the produced 
gluon density distribution, computed with the publicly available fKLN code 
\cite{fKLN}. In central Au+Au collisions, both profiles are normalized to 
the same total entropy (adjusted to reproduce the total final charged hadron 
multiplicity $dN_\mathrm{ch}/dy$ in these collisions) and converted to 
energy density using the equation of state s95p-PCE (see next section).
With this normalization, both initializations correctly describe the 
centrality dependence of $dN_\mathrm{ch}/dy$ for {\it ideal} fluid dynamics 
(i.e. for isentropic expansion). 

In the viscous case, viscous heating produces additional entropy, 
resulting in larger final multiplicities which we must correct 
for by renormalizing the initial entropy density profile in such a way
that the final multiplicity is held fixed. We perform this renormalization 
for the 5\% most central Au+Au collisions (i.e.\ at $b\eq2.33$\,fm) and then
keep the resulting normalization constant fixed for non-central collisions,
i.e. we again assume that the models produce the correct dependence of 
initial entropy production on collision geometry. It is known, however,
that the fractional increase of the final entropy over its initial value 
due to viscous heating depends on the size of the collision fireball
\cite{Song:2008si} and is therefore expected to be larger in peripheral 
than central Au+Au collisions. For the results presented in this paper,
we have checked that the centrality dependence of viscous entropy 
production is sufficiently weak so that it does not strongly modify the 
centrality dependence of $dN_\mathrm{ch}/dy$.  

%

Figure \ref{F1} shows that the energy density profile from the CGC 
initialization has a steeper surface gradient than the Glauber profile.
This leads to larger radial acceleration (i.e. radial flow develops
more quickly) and is also in part responsible for the larger spatial
eccentricity of the CGC profiles at nonzero impact parameters when 
compared to the Glauber eccentricities \cite{Hirano:2005xf,Heinz:2009cv}.

The shear viscous pressure tensor $\pi^{\mu\nu}$ is initialized
with its Navier-Stokes value $\pi^{\mu\nu}\eq2\eta\sigma_{0}^{\mu\nu}$
where $\sigma_{0}^{\mu\nu}$ is the velocity shear tensor at time
$\tau_0$, calculated from the initial Bjorken velocity profile, $u^\mu\eq
(u^\tau,u^x,u^y,u^\eta)\eq(1,0,0,0)$.

For the medium's viscous properties the shear viscosity $\eta/s$ is the 
key parameter. According to perturbative and lattice QCD, the temperature 
dependence of $\eta/s$ is weak over the range explored in heavy-ion 
collisions at RHIC energies. This suggests use of a constant ratio 
$\eta/s$. In this work, the value of $\eta/s$ is tuned from 0.08 to 0.24 in
order to study the effects of shear viscosity on the hadron spectra and 
elliptic flow. The influence of a temperature dependent $\eta/s$ will be
explored in a forthcoming paper~\cite{ChunShen2010}. 

Final state hadron spectra are calculated from the hydrodynamics 
output via the Cooper-Frye procedure \cite{Cooper:1974mv}
\begin{eqnarray}
\label{Cooper}
  E\frac{d^3N_i}{d^3p} &=& \frac{g_i}{(2\pi)^3}\int_\Sigma 
  p\cdot d^3\sigma(x)\, f_i(x,p),
\end{eqnarray}
where $\Sigma$ is the freeze-out surface with normal vector 
$d^3\sigma_\mu(x)$. We take for $\Sigma$ an isothermal surface; 
calculations for different freeze-out temperatures are presented 
in Sec.~\ref{sec4b}. After computing the spectra of all hadronic 
resonances included in EoS s95p-PCE from Eq.~(\ref{Cooper}), we use the
resonance decay program \cite{Sollfrank:1990qz,Sollfrank:1991xm} from 
the {\tt AZHYDRO} package~\footnote{{\tt AZHYDRO} is available at the URL\\
{\tt http://www.physics.ohio-state.edu/\~{}froderma/}.} to let the unstable 
resonances decay. The pion and proton spectra shown in this work include all 
decay products from strong decays.

The distribution function on the freeze-out surface can be decomposed as 
$f\eq{f}_\mathrm{eq}+\delta f$ into a 
local equilibrium part
\begin{eqnarray}
\label{f0}
  f_\mathrm{eq}(p,x) 
  = \frac{1}{e^{p\cdot u(x)/T(x)}\pm 1}
\end{eqnarray}
and a (small) deviation $\delta f$ from local equilibrium due to shear 
viscous effects for which we make the quadratic ansatz 
\cite{Teaney:2003kp,Baier:2006um} (for other possibilities see
\cite{Dusling:2009df}) using
\begin{eqnarray}
\label{deltaf}
  \!\!\!\!
  \delta f(x,p)\!\!&=&\!\! 
  f_\mathrm{eq}(p,x) \bigl(1{\mp}f_\mathrm{eq}(p,x)\bigr)
  \frac{p^\mu p^\nu \pi_{\mu\nu}(x)}{2T^2(x)\left(e(x){+}p(x)\right)}
\end{eqnarray}
(the upper (lower) sign is for fermions (bosons)) for all particle species. 
$\delta f$ is proportional to the shear viscous pressure tensor $\pi^{\mu\nu}(x)$ 
on the freeze-out surface and increases (in our case) quadratically with
the particle momentum.

\section{Equation of State}
\label{sec3}

To solve the equations (\ref{ISeq1}) and (\ref{ISeq2}) one has to know the 
equation of state $p(e)$ (EoS) of the medium. In this work we compare
three different equations of state,
in order to study how the EoS affects the hadron spectra and elliptic flow. 
Two of them, SM-EOS~Q \cite{Song:2007ux} and EOS~L \cite{Song:2008si} are 
well known in the literature; the former implements a (slightly smoothed) 
first order phase transition between an ideal massless parton gas and a 
hadron resonance gas (HRG), the second is a rough attempt to match lattice 
QCD (LQCD) data \cite{Aoki:2005vt} above $T_\mathrm{c}$ to the HRG in a 
smooth crossover transition, as seen in LQCD. In both cases, the system is 
assumed to be in chemical equilibrium all the way down to kinetic freeze-out 
at temperature $T_\mathrm{dec}$.

Our third equation of state, s95p-PCE, also interpolates between the HRG
at low temperature and the lattice EoS at high temperatures, but the
matching procedure is more sophisticated than the one used to construct 
EOS~L, and the lattice EoS is based on the recent results by the hotQCD 
collaboration~\cite{Cheng:2007jq,Bazavov:2009zn}. Furthermore, below 
$T_\mathrm{chem}\eq165$\,MeV temperature, the EoS is that of a 
\emph{chemically frozen} HRG. The matching procedure using a chemically 
equilibrated HRG is explained in detail in~\cite{Huovinen:2009yb}. The 
procedure for the chemically frozen HRG is identical since the chemical 
freeze-out temperature is below the temperature where the interpolated 
EoS deviates from the HRG EoS.

\begin{figure}[ht]
\includegraphics[width=\linewidth,clip=]{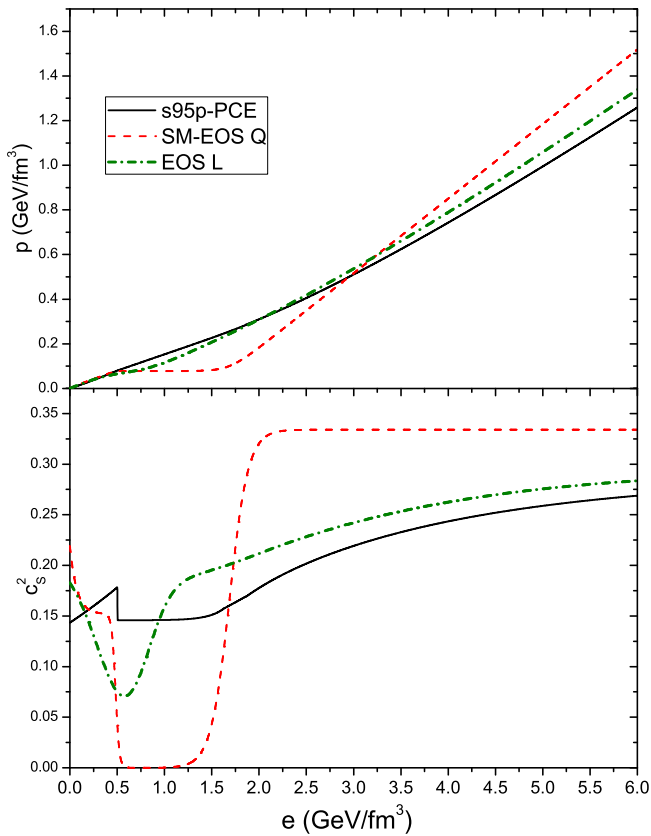}
\caption{(Color online) The three equations of state s95p-PCE, SM-EOS~Q, 
and EOS~L used in this paper. The lower panel shows the squared speed of 
sound $c_{s}^2 = \frac{\partial p}{\partial e}$ as a function of 
energy density $e$ whereas $p(e)$ is shown in the upper panel.
}
\label{F2}
\end{figure}
%

However, the version of s95p-PCE used here deviates slightly from the 
s95p-PCE-v1 EoS shown in Appendix C of Ref.~\cite{Huovinen:2009yb}. First, 
we have chosen $T_\mathrm{chem}\eq165$\,MeV for the chemical freeze-out 
temperature, as fitted to experimental data using thermal models
\cite{Adams:2005dq,BraunMunzinger:2001ip,Andronic:2008gu}, and we have 
considered as stable particles those with half-life larger than 40 fm/$c$ 
instead of 10 fm/$c$. Second, our s95p-PCE corresponds to a historically
slightly earlier stage of the parametrization of the EoS than the final 
version published in Ref.~\cite{Huovinen:2009yb}: The fit to the lattice 
data was done without the $T=630$\,MeV data point. This causes at most 
0.4\% difference between this version and the final version of the EoS. We 
have checked that such a small difference does not cause observable 
consequences in the fluid-dynamical evolution~\footnote{For a discussion 
of the uncertainties in parametrizing the lattice data and its effect on
fluid dynamics see~\cite{Huovinen:2009yb}.}.

We have built the EoS of the chemically frozen hadron gas using the
standard procedure in the literature: Below $T_\mathrm{chem}$ the
ratios of stable hadron yields are fixed to their chemical equilibrium
values at $T_\mathrm{chem}$ by finite non-equilibrium chemical
potentials $\mu_i(T)$
\cite{Bebie:1991ij,Hirano:2002ds,Kolb:2002ve,Teaney:2002aj,Huovinen:2007xh}.
It is worth noting that the ratios of individual particle densities
are not conserved. What is conserved are the ratios of the {\em total 
densities} of stable particles, $\bar{n}_i$, where total density means 
the sum of the actual density of species $i$ and the additional density 
of the same species that would arise if all unstable resonances in the 
system were allowed to immediately and irreversibly decay. The rapid 
processes that form and decay resonances through strong interactions 
are still in equilibrium, and thus the resonance populations are in 
equilibrium with the populations of their daughter particles (see 
\cite{Bebie:1991ij,Hirano:2002ds} for a detailed discussion). Thus 
the chemically frozen system is in a state of partial chemical 
equilibrium (PCE).

In practice the chemically frozen EoS is evaluated assuming that the 
evolution is isentropic and the ratios $\bar{n}_i/s$ stay constant. Strictly
speaking this is not the case in viscous hydrodynamics since dissipation 
causes an increase in entropy. However, we have checked that in our
calculations the viscous entropy production from fluid cells with
temperatures below $T_\mathrm{chem}\eq165$\,MeV is small (see also the 
right panel of Fig.~8 in \cite{Song:2008si}) and our EoS is a good 
approximation of the physical EoS.

An analytic parametrization of s95p-PCE is given in Appendix~\ref{appa}; 
the EoS can be obtained in a tabulated form at~\cite{s95p-PCE}, where the 
particles included in the hadron resonance gas are also listed. (We included
all resonances listed in the summary of the 2004 edition of the Review of 
Particle Physics~\cite{PDG2004} up to 2 GeV mass. Note that our s95p-PCE is 
called s95p-PCE165-v0 at \cite{s95p-PCE}, to differentiate it from other 
versions of the parametrization.)

The three equations of state are compared in Fig.~\ref{F2}. The upper panel
shows the pressure and the lower panel the squared speed of sound as a function 
of $e$. The spike in $c_s^2(e)$ at $e\sim0.5$\,GeV/fm$^3$ results from the 
sudden breaking of chemical equilibrium at $T_\mathrm{chem}\eq165$\,MeV. It 
has negligible consequences for the expansion dynamics. Fig.~\ref{F2} shows 
that s95p-PCE is a much softer EoS than SM-EOS~Q in the QGP phase above 
$T_\mathrm{c}$ but much harder in the phase transition region around 
$T_\mathrm{c}$. Contrary to SM-EOS~Q and EOS~L, the rapid crossover 
transition between quarks and hadrons that is realized by nature doesn't 
have a well-defined ``softest point'' \cite{Hung:1994eq} which would cause 
the fireball to spend an extended time period in the critical region. 
Instead, the speed of sound never drops much below its value in the HRG, 
causing the fireball to cool rapidly through the phase transition 
\cite{Zschiesche:2001dx}.

\section{Spectra and elliptic flow}
\label{sec4}

In this section, we discuss the dependence of the transverse momentum
spectra in central $200\,A$\,GeV Au+Au collisions (0-5\% centrality,
$b\eq2.33$\,fm) and the elliptic flow $v_2(p_T)$ in semiperipheral
collisions (20-30\% centrality, $b\eq 7.5$\,fm) for pions, protons and
(for $v_2$) all charged hadrons, on the EoS and various input parameters 
discussed in Secs.~\ref{sec2} and \ref{sec3}. We have also checked that 
everything we say below about the central collision spectra also applies, 
at the same level of precision, to the $\phi$-averaged spectra in 
semiperipheral collisions.

Since the amount of viscous heating depends on the input parameters, we 
retune for each case the normalization of the initial energy density profile 
{\em in central collisions} so that the same final $\pi^+$ multiplicity 
density $dN_{\pi^+}/dy$ is obtained. Its value is adjusted by eye such 
that an optimal fit to the measured pion spectrum is obtained in the 
low-$p_T$ region, $p_T<1.5$\,GeV/$c$. As there are slight discrepancies 
between the published data from the STAR and PHENIX Collaborations, and 
these experiments give their results in different centrality bins, we have 
decided to concentrate on PHENIX results \cite{Adare:2010ux,Adler:2003cb} 
when comparing the theoretical curves with experimental data. Since we do 
not attempt to fit these data but use the comparison only to illustrate 
trends, this procedure is acceptable. A future serious dynamical model fit 
to the data will require proper accounting for systematic uncertainties 
and discrepancies between the different experiments.  

Since viscous heating effects are relatively more important in peripheral
than in central collisions, our renormalization to constant multiplicities 
at $b\eq2.33$\,fm leads to slightly different pion multiplicities at 
larger impact parameters. For a given EoS, ensuring the same final pion 
multiplicity is equivalent to ensuring the same final total multiplicity. 
For different equations of state (see Sec.~\ref{sec4d}) identical pion 
multiplicities correspond to slightly different total multiplicities.

In the following we show hadron spectra and elliptic flow up to transverse 
momenta of 3\,GeV/$c$. We emphasize that this is for illustrative purposes
only and does not imply that we believe hydrodynamics to provide a valid 
description up to such large $p_T$. When comparing model results with 
experimental data, we judge the quality of agreement by focussing on the 
region $p_T<1.5$\,GeV/$c$ for pions and $p_T<2.5$\,GeV/$c$ for protons (which 
is where we believe hydrodynamics is a reliable approach \cite{Heinz:2004pj}). 
Specifically for pions, if the calculated spectra drop off more steeply
than the measured ones above $p_T\eq1.5$\,GeV/$c$, we discount this 
discrepancy, noting that this is the region where the experimental spectra 
begin to change from an exponential to a power-law shape due to the onset 
of hard physics.
  
\subsection{$\bm{\eta/s}$-dependence at fixed 
$\bm{\tau_0 = 0.4}$\,fm/$\bm{c}$ and $\bm{T_\mathrm{dec} = 140}$\,MeV}
\label{sec4a}

\begin{figure}[thb]
\includegraphics[width=1\linewidth,clip=]{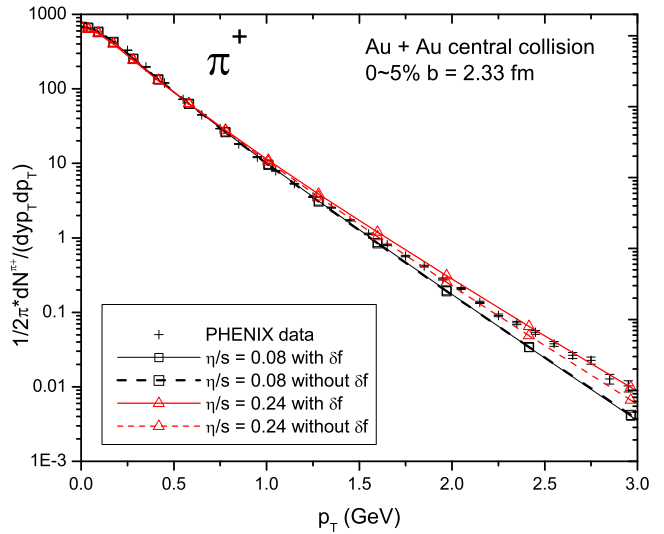}
\caption{(Color online) Pion spectra for $200\,A$ GeV Au+Au collisions at 
0-5\% centrality from {\tt VISH2+1}, compared with PHENIX data 
\cite{Adler:2003cb}. Results for two different constant values of $\eta/s$ 
(0.08 and 0.24) are shown; strong resonance decays are included. Solid and 
dashed lines show the spectra calculated from the full distribution function 
$f\eq{f}_\mathrm{eq}+\delta f$ (``with $\delta f$'') and from the 
equilibrium part only (``without $\delta f$''). The hydrodynamic 
evolution starts at $\tau_0\eq0.4$\,fm/$c$ with an initial CGC energy 
density profile and ends at $T_\mathrm{dec}\eq140$\,MeV. The EoS is s95p-PCE.
}
\label{F3}
\end{figure}

\begin{figure*}[htb]
\includegraphics[width=1.0\linewidth,clip=]{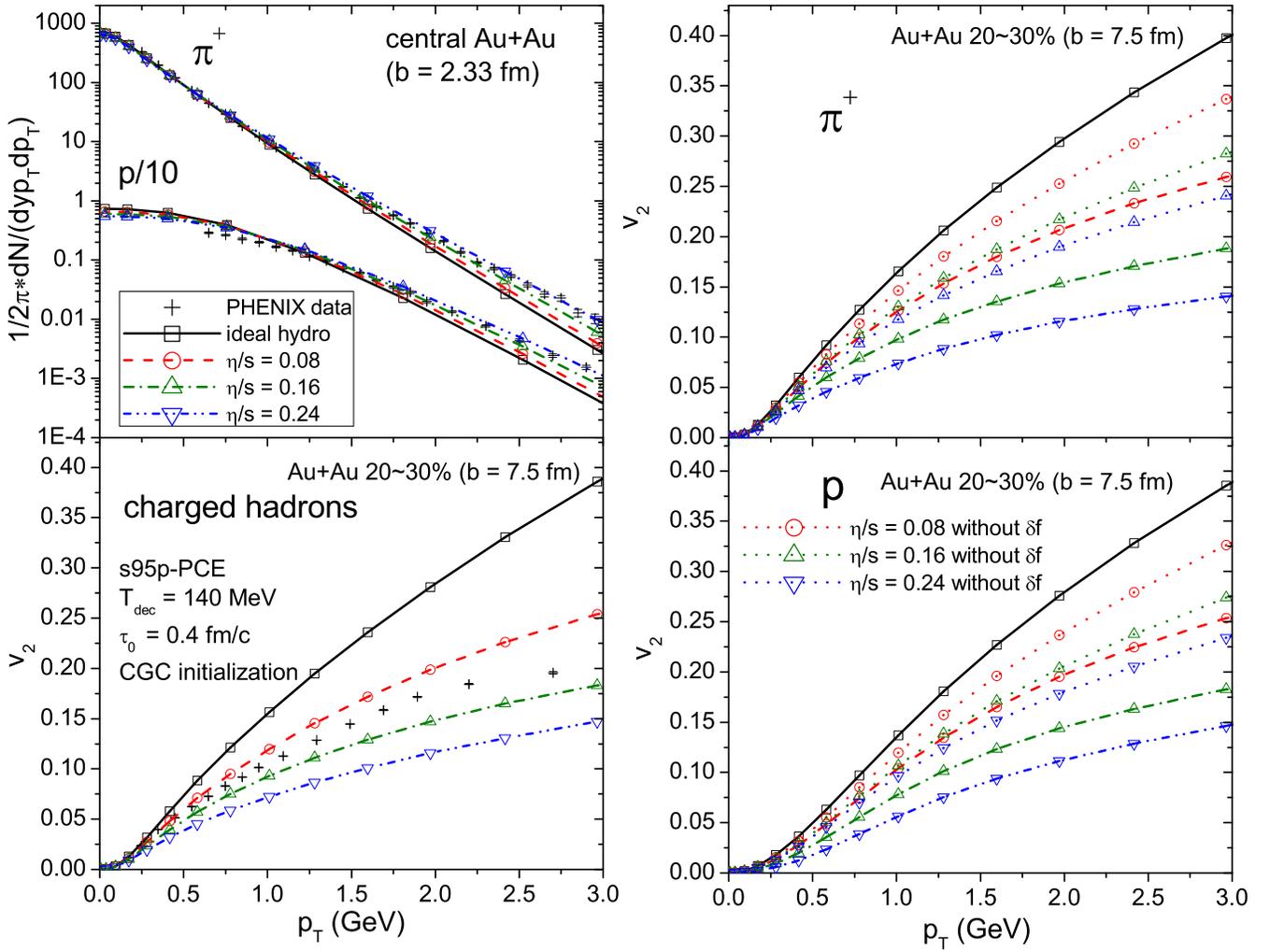}
\caption{{\sl Upper left panel:} Transverse momentum spectra $dN/(dy\,d^2p_T)$ 
for pions and protons from {\tt VISH2+1} for the 5\% most central Au+Au 
collisions ($b\eq2.33$\,fm), compared with experimental data from the PHENIX 
Collaboration \cite{Adler:2003cb}. {\sl Lower left panel:} Differential 
elliptic flow $v_2(p_T)$ for charged hadrons from Au+Au collisions at 20-30\% 
centrality ($b\eq7.5$\,fm), compared with PHENIX data \cite{Adare:2010ux}.
{\sl Right panels:} $v_2(p_T)$ for pions (top) and protons (bottom).
In these panels we compare the elliptic flow computed from the full 
distribution function $f\eq{f}_\mathrm{eq}+\delta f$ (dashed, dot-dashed
and double-dot-dashed lines) with the contribution from the equilibrium part 
only (dotted lines, ``without $\delta f$''). Lines with different symbols 
show calculations for different constant values of the specific shear 
viscosity $\eta/s$, ranging from 0 (ideal hydro, solid lines) to 0.24 as 
indicated. All strong resonance decays are included; charged hadrons comprise 
$\pi^{\pm}$, $K^{\pm}$, p, $\overline{p}$, $\Sigma^{\pm}$, 
$\overline{\Sigma}^{\mp}$, $\Xi^{-}$, $\overline{\Xi}^{+}$, $\Omega^{-}$, 
and $\overline{\Omega}^{+}$. The EoS, initial and final conditions are the 
same as in Fig.~\ref{F3}.
}
\label{F4}
\end{figure*}

Transverse momentum spectra of pions and protons in the most central 
Au+Au collision are shown in Fig.~\ref{F3} and in the upper left panel 
of Fig.~\ref{F4}. The spectra include all strong resonance decays. Here 
we hold initial and final conditions fixed (except for a renormalization 
of the initial peak energy density to ensure the same final multiplicity 
in all calculations) and vary the specific shear viscosity $\eta/s$ (see 
figure captions for details). One sees that under these conditions larger
$\eta/s$-values result in flatter spectra; the effect is particularly 
strong for protons at low $p_T$. The main reason is that larger shear 
viscosity leads to larger radial flow, due to a positive contribution 
from $\pi^{\mu\nu}$ to the effective transverse pressure gradients at 
early times \cite{Teaney:2004qa,Chaudhuri:2005ea,Heinz:2005bw}. 

Figure~\ref{F3} identifies, however, a second contribution to the viscous 
hardening of the spectra: for $\eta/s\eq0.16$ and 0.24 and evolution with 
s95p-PCE, we find that the viscous correction due to the non-equilibrium 
deviation $\delta f$ of the distribution function on the freeze-out surface, 
Eq.~(\ref{deltaf}), is {\em positive} for $p_T{\,\agt\,}0.5-1$\,GeV/$c$
\footnote{The Landau matching conditions require the $\delta f$ 
correction to integrate to zero when summing over all momenta, so a 
positive $\delta f$ contribution at high $p_T$ implies a negative 
$\delta f$ contribution at low and/or intermediate $p_T$. In 
\cite{Song:2007ux} we found that it typically changes sign twice.},
thus adding to the hardening of the spectra from radial flow
\footnote{We checked that for all equations of state studied here
that the sign of $\delta f$ at high $p_T$ does not depend on whether
we use CGC or Glauber initial conditions.}.
This is the same sign for $\delta f$ as found in \cite{Dusling:2007gi} 
(for a different EoS) but opposite to what had been found earlier with 
{\tt VISH2+1} for smaller values of $\eta/s$ using SM-EOS~Q (i.e. a first 
order phase transition) \cite{Song:2007ux}. (For $\eta/s\eq0.08$ 
Fig.~\ref{F3} shows a {\em negative} $\delta f$ correction for pions at 
large $p_T$, of {\em same} sign but much smaller magnitude than found 
earlier \cite{Song:2007ux} with Glauber initial conditions and SM-EOS~Q).
Our finding confirms the fragility of the sign of $\delta f$ that was 
already discussed in \cite{Song:2007ux}
\footnote{For EOS~L and SM-EOS~Q and $\eta/s>0.08$, we find a 
negative sign of the $\delta f$ contribution to both pion and proton 
spectra at high $p_T$, while the corresponding contribution is positive
in the case of s95p-PCE. The negative sign appears to be correlated with 
the use of an EoS with a ``softest point''. From ideal fluid dynamic 
simulations with such first-order or almost-first-order phase transitions 
we know that the rapid change of $c_s^2$ in the transition region generates 
strong structures in the radial velocity profile in fireball regions that 
are close to the critical temperature \cite{Kolb:2000sd}, and that these 
structures partially survive until the matter has reached decoupling. We 
suspect that velocity gradients associated with these structures play an 
important role in generating for EOS~L and SM-EOS~Q a negative $\delta f$ 
contribution to the spectra at high $p_T$.}.

We note in passing that the positive $\delta f$ at large $p_T$ found 
here with s95p-PCE is found to be largest in near-central collisions 
($b{\,\approx\,}0$) where it can even lead to a {\em positive} 
$\delta f$-correction to the differential elliptic flow $v_2(p_T)$. At 
larger $b$, the $\delta f$-contribution to $v_2(p_T)$ remains negative 
here (see right panels of Fig.~\ref{F4}), as has been consistently 
observed in other work \cite{Romatschke:2007mq,Song:2007fn,Song:2007ux,%
Dusling:2007gi,Luzum:2008cw}.

%
%

%
%

The time-integrated effect of the shear viscous pressure on the radial 
flow and the ``instantaneous'' effect of the viscous correction $\delta f$
to the distribution function on the freeze-out surface together give the 
total shear viscous correction to the hadron spectra. For $\eta/s\eq0.08$ 
we see in Fig.~\ref{F3} that with s95p-PCE the $\delta f$ correction to 
the pion spectrum is almost negligible, but the upper left panel in 
Fig.~\ref{F4} shows that the pion and proton spectra are still flatter 
than for the ideal fluid, reflecting the larger radial flow caused by 
the shear viscous increase of the transverse pressure gradients 
\cite{Song:2007ux}. Thus both the effect of viscosity on radial flow 
{\em and} $\delta f$ contribute to the flattening of the hadron spectra.

Comparing with the experimental data we find that both pion and 
proton spectra favor a relatively large shear viscosity, 
$\eta/s = 0.16 \sim 0.24$. We caution that this conclusion is based
on calculations done with constant (i.e. temperature independent)
$\eta/s$ and may be subject to revision once one properly accounts 
for the increase of $\eta/s$ in the dilute late hadronic stage.

Proceeding to the elliptic flow, we start with a discussion of the charged
hadron $v_2$ in the lower left panel of Fig.~\ref{F4}. Here, larger shear
viscosity values are seen to lead to a stronger suppression of elliptic 
flow. The right panels in Fig.~\ref{F4} show that this suppression
is again the consequence of two additive effects: shear viscosity reduces
the buildup of anisotropic collective flow, reflected in the 
equilibrium part $f_\mathrm{eq}$ of the distribution function on the
freeze-out surface (dotted lines in Fig.~\ref{F4}), but the viscous 
correction $\delta f$ causes an additional suppression of $v_2$. For 
$T$-independent $\eta/s$, both suppression effects increase monotonically 
with shear viscosity; however, the increase of the $\delta f$-correction 
with rising $\eta/s$ is weaker than that of the viscous suppression 
of the collective flow anisotropy. The stronger suppression of $v_2$
for larger $\eta/s$ is thus mostly due to the viscous suppression of
anisotropic flow.    

Since elliptic flow data for identified pions and protons 
in the particular centrality bin shown in Fig.~\ref{F4} are not yet 
available, we compare in the lower left panel with experimental data for 
unidentified charged hadrons. This plot suggests that, even for CGC initial
conditions which produce more eccentric fireballs than the Glauber model
\cite{Hirano:2005xf,Luzum:2008cw}, the $v_2$ data suggest a smaller value
%
\begin{figure*}[hbt]
\includegraphics[width=\linewidth,clip=]{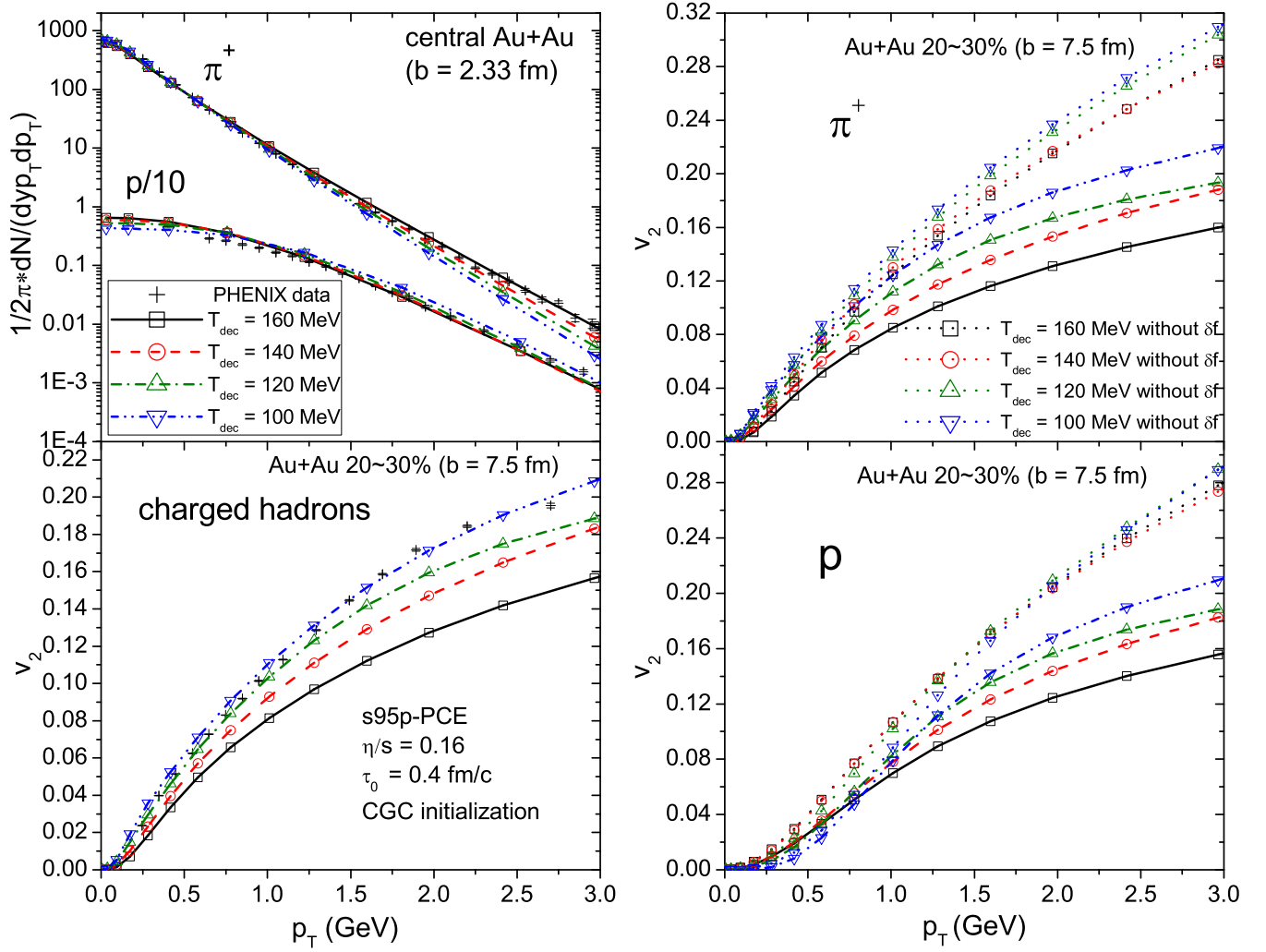}
\caption{Similar to Fig.~\ref{F4}, but for fixed $\eta/s\eq0.16$ and
varying decoupling temperature $T_\mathrm{dec}$ ranging from
100 to 160\,MeV as indicated.
}
\label{F5}
\end{figure*}
%
for $\eta/s$, $\eta/s = 0.08 - 0.16$, than obtained from the $p_T$-spectra 
for central collisions. 

This tension between the slope of the $p_T$-spectra (which tends to favor 
larger $\eta/s$ values) and the $p_T$-dependence of $v_2$ (which favors 
smaller values) is generic and, as far as we were able to ascertain, cannot 
be resolved with purely hydrodynamic calculations that assume constant 
$\eta/s$. A possible solution of this problem will likely involve accounting
for temperature-dependence of $\eta/s$ and/or the transition to a microscopic
kinetic description for the late hadronic stage. 

\subsection{$\bm{T_\mathrm{dec}}$-dependence at fixed 
$\bm{\tau_0 = 0.4}$\,fm/$\bm{c}$ and $\bm{\eta/s\eq0.16}$}
\label{sec4b}

In Fig.~\ref{F5} we explore the sensitivity of spectra and elliptic flow
on the value of the decoupling temperature, holding all other parameters 
fixed. For the constant $\eta/s$ we select $\eta/s\eq0.16$ as a compromise
between the values preferred by the proton spectra and charged hadron $v_2$,
respectively, in Fig.~\ref{F4}. 

The left upper panel shows that lower freeze-out temperatures lead to
flatter proton spectra. This is a consequence of additional radial flow
built up during the extra time the fireball needs to cool down to lower
$T_\mathrm{dec}$. As is well known \cite{Schnedermann:1992ra,%
Schnedermann:1993ws}, the heavier protons receive a larger push to higher
$p_T$ from radial flow than the lighter pions. Indeed, Fig.~\ref{F4}
shows that the pion spectra become {\em steeper} as $T_\mathrm{dec}$ is 
lowered \cite{Hirano:2005wx}. Since pions are almost massless on the
scale of measured transverse momenta, the inverse slope of their 
$p_T$-spectrum can be approximated by the relativistic blueshift
formula \cite{Schnedermann:1992ra,Schnedermann:1993ws}
$T_\mathrm{slope}\eq{T}_\mathrm{dec}\sqrt{\frac{1+\langle v_{\perp}\rangle}
{1-\langle v_{\perp}\rangle}}$ where $\langle v_{\perp}\rangle$ is the 
average radial flow at $T_\mathrm{dec}$. For pions, the steepening 
effects on their spectrum from decreasing $T_\mathrm{dec}$ overwhelm
the flattening effects resulting from the associated increase of 
$\langle v_{\perp}\rangle$, causing a net softening of the pion 
spectra for lower freeze-out temperatures.

>From the lower left panel of Fig.~\ref{F5} one sees that lower
decoupling temperatures lead to larger elliptic flow $v_2(p_T)$ for
charged hadrons. To fully understand this systematics it is worth 
comparing charged hadrons to the $p_T$-spectra and $v_2(p_T)$ of pions 
(upper left and right panels, respectively) which dominate the charged 
hadron yield. The observed tendency reflects a combination of three effects:

(i) Since the $p_T$-spectrum of pions (which dominate the charged hadrons) 
gets steeper, even the same hydrodynamic momentum anisotropy would lead to 
a larger slope of $v_2(p_T)$, to compensate for the lower yield at high $p_T$. 

(ii) Since the fireball hasn't lost all of its eccentricity by the time the 
QGP converts to hadrons \cite{Song:2007ux}, additional momentum anisotropy 
is generated during the hadronic stage. Lower decoupling temperatures give 
the system time to develop more momentum anisotropy, leading to a larger 
$v_2$. If the $p_T$-spectrum stays unchanged or gets steeper (as is the
case for pions in Fig.~\ref{F5}), a larger $v_2$ must lead to a larger 
$v_2(p_T)$. The combination of effects (i) and (ii) is seen in the dotted 
lines in the upper right panel, which reflect the hydrodynamic flow 
anisotropy at decoupling, undistorted by viscous corrections $\delta f$ to 
the local equilibrium distributions at freeze-out. The effect (ii) decreases 
with increasing $\eta/s$ in the hadronic phase (not shown here), so the 
combined effect may be weaker than seen in Fig.~\ref{F5} if viscous 
hydrodynamics is replaced by a microscopic hadron cascade such as UrQMD in 
the hadronic phase. 

(iii) The (negative) viscous corrections from $\delta f$ to $v_2$ are 
smaller at lower temperatures, due to the general decrease of the
viscous pressure components \cite{Song:2007ux}. This contributes the
largest fraction of the observed increase of $v_2(p_T)$ with decreasing 
$T_\mathrm{dec}$, especially at large $p_T$.

Combining the information from the two left panels in Fig.~\ref{F5}
we conclude that both the proton spectra in central collisions and 
charged hadron $v_2(p_T)$ in peripheral collisions favor decoupling 
temperatures near the lower end of the window studied here (i.e.
$T_\mathrm{dec}\eq100$\,MeV works better than $T_\mathrm{dec}\eq140$\,MeV).
The pion spectra are affected by variations of $T_\mathrm{dec}$ mostly
at $p_T{\,\agt\,}1-1.5$\,GeV/$c$ where they fall increasingly below the
experimental data as we lower $T_\mathrm{dec}$. However, this is also the
region where the hydrodynamic description of the pion spectra is known to 
begin to break down \cite{Heinz:2004pj}, due to the gradual transition from
soft to hard physics which causes the pion spectrum to change from an 
exponential to a power-law shape. Focusing therefore on the region 
$p_T{\,<\,}1.5(2.5)$\,GeV/$c$ for pions (protons), we conclude that a
purely hydrodynamic description of the experimental data favors 
freeze-out temperatures near 100\,MeV.

The right panels of Fig.~\ref{F5} show how $T_\mathrm{dec}$ affects the 
elliptic flow of different identified hadrons. Charged hadrons mostly 
reflect the behavior of the dominating pions whose $v_2(p_T)$ increases 
with decreasing freeze-out temperature. But protons behave differently:
At low $p_T{\,<\,}1$\,GeV, their elliptic flow decreases with decreasing
decoupling temperature, while at high $p_T$ it increases with decreasing
$T_\mathrm{dec}$. The latter feature reflects the increasing hydrodynamic 
momentum anisotropy and decreasing magnitude of the $\delta f$ correction, 
just like it is reflected in the pion and charged hadron $v_2$. The 
decrease of proton $v_2$ at low $p_T$, on the other hand, is a consequence
of having larger radial flow at lower $T_\mathrm{dec}$ which pushes the
protons to larger $p_T$. So, rather than thinking of this effect as a
decrease of proton $v_2$ at fixed $p_T$, we should think of it as shifting
the elliptic flow to larger $p_T$.

\begin{figure*}
\includegraphics[width=\linewidth,clip=]{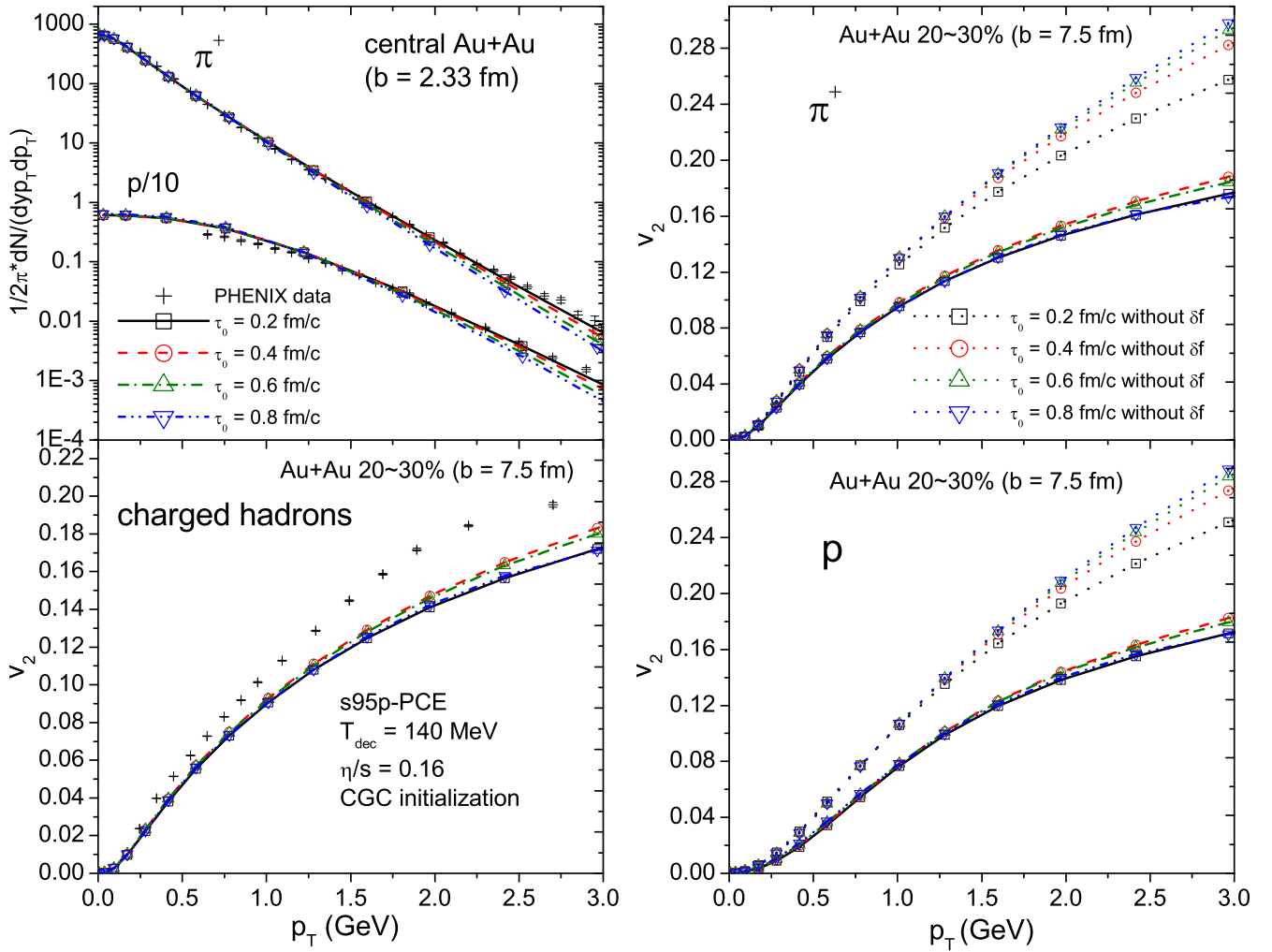}
\caption{Similar to Fig.~\ref{F4}, but for fixed $\eta/s\eq0.16$ and
varying starting time $\tau_0$ for the hydrodynamic evolution,
ranging from 0.2 to 0.8\,fm/$c$ as indicated.
}
\label{F6}
\end{figure*}
%

\subsection{$\bm{\tau_0}$-dependence at fixed $\bm{\eta/s\eq0.16}$ 
and $\bm{T_\mathrm{dec}\eq140}$\,MeV}
\label{sec4c}

The upper left panel of Fig.~\ref{F6} shows that the pion and proton 
spectra react similarly to a change of the starting time $\tau_0$
of the hydrodynamic evolution: smaller $\tau_0$ values lead to more
high-$p_T$ particles, reflecting more radial flow. Starting hydrodynamics 
earlier allows it to generate radial flow earlier, and even though this 
also causes the fireball to cool down to $T_\mathrm{dec}$ sooner and 
freeze out earlier, the net effect is still a slight increase of the 
average radial flow at freeze-out.

For soft momenta $p_T{\,<\,}1.5$\,GeV/$c$, the effect of $\tau_0$
on $v_2(p_T)$ is negligible. This is true even for protons, showing
that the increase of radial flow with decreasing $\tau_0$ is a small
effect and not enough to visibly push the proton $v_2$ to larger $p_T$. 
At higher $p_T$, the dependence of the charged hadron, pion and proton 
$v_2$ on $\tau_0$ is non-monotonic. The right panels of Fig.~\ref{F6} 
show that this non-monotonic behaviour is the result of two 
counteracting tendencies which both depend on $\tau_0$ monotonically:
(i) The elliptic flow computed from the local equilibrium part $f_\mathrm{eq}$
of the distribution function at freeze-out increases monotonically
with increasing $\tau_0$, reflecting the longer total fireball lifetime 
(and thus the longer time available to build up momentum anisotropy) when
the hydrodynamic evolution starts later. (ii) The $v_2$-suppression 
resulting from the viscous correction $\delta f$ at freeze-out also 
increases monotonically with increasing $\tau_0$. We don't have a complete
understanding of why starting (and thus also ending) the hydrodynamics 
later leads to a larger $\delta f$ on the decoupling surface; we suspect 
that since the hydrodynamical flow would eventually settle into a
three-dimensional spherically symmetric Hubble flow with no shear
stress, starting earlier leads to a stronger transverse flow, and thus
to a flow profile which is closer to a spherically symmetric flow at
the time of decoupling.

\subsection{EoS dependence at fixed $\bm{\tau_0\eq0.4}$\,fm/$\bm{c}$, 
$\bm{\eta/s\eq0.16}$, and $\bm{T_\mathrm{dec}\eq140}$\,MeV}
\label{sec4d}

\begin{figure*}
\includegraphics[width=\linewidth,clip=]{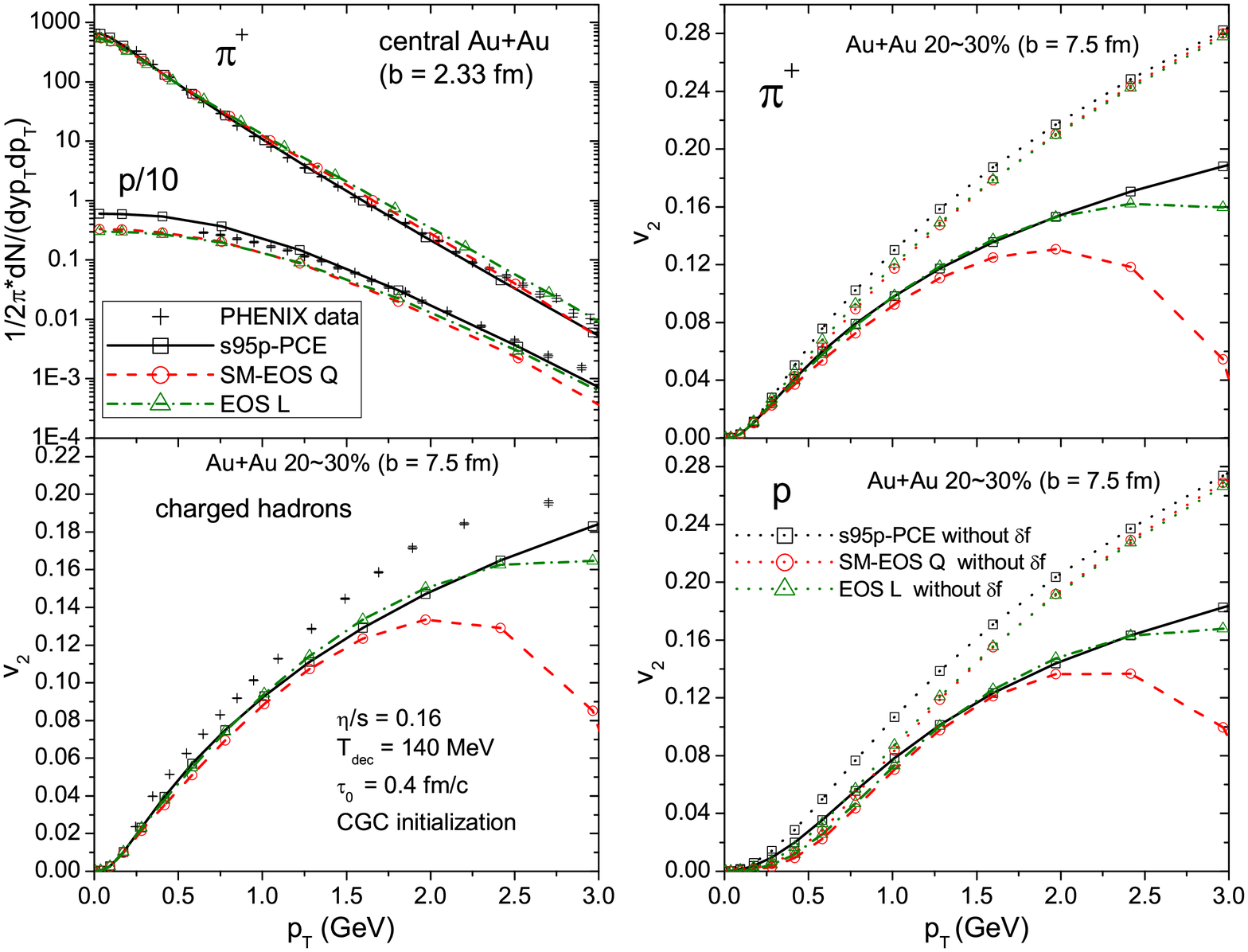}
\caption{Similar to Fig.~\ref{F4}, but for fixed $\eta/s\eq0.16$ and
different equations of state (SM-EOS~Q, EOS~L, and s95p-PCE) as 
indicated. Since SM-EOS~Q and EOS~L have different chemical composition
than s95p-PCE at $T_\mathrm{dec}\eq140$\,MeV, they yield fewer protons
than s95p-PCE when normalized to the same pion yield.
}
\label{F7}
\end{figure*}
%

In Fig.~\ref{F7} we study the sensitivity of hadron spectra and elliptic 
flow on the equation of state, holding all other hydrodynamic parameters 
fixed (except for the normalization of the initial energy density profile 
which is again adjusted to ensure constant final multiplicity in central 
Au+Au collisions). We first note that, due to the different chemical 
composition at hadron freeze-out, the proton yields for EOS~L and SM-EOS~Q 
are below those of s95p-PCE if we hold the pion mulitiplicity fixed: In 
s95p-PCE we prohibit protons from annihilating on antibaryons while such 
annihilation processes are allowed in the other two equations of state 
which assume hadrons in chemical equilibrium. To explore flow effects we 
should concentrate on the {\em shape} (i.e. inverse slopes) of the pion 
and proton spectra. We see that EOS~L produces the flattest spectra, 
followed by SM-EOS~Q, whereas the spectra from s95p-PCE are steepest. Since 
all three curves correspond to the same (constant) freeze-out temperature 
$T_\mathrm{dec}\eq140$\,MeV, these differences can only arise from different
amounts of radial flow or different $\delta f$ corrections (i.e. different
viscous pressure components $\pi^{\mu\nu}$) along the freeze-out surface.
To separate these two effects we plotted the spectra calculated without the
$\delta f$ correction and found the same hierarchy. We conclude that, for
fixed freeze-out temperature, s95p-PCE produces the weakest radial flow
averaged over the freeze-out surface, EOS~L generates the strongest flow,
with SM-EOS~Q falling in between.

The reasons for s95p-PCE generating less radial flow than the other two
equations of state are complex and subtle. The differences in speed of 
sound during the evolution largely cancel out
(see Ref.~\cite{Huovinen:2009yb}). 
The key difference is that, at a fixed freeze-out temperature, the
chemically frozen HRG embodied in s95p-PCE has a considerably larger
energy density ($e_\mathrm{dec}\eq0.301$\,GeV/fm$^3$ at 
$T_\mathrm{dec}\eq140$\,MeV) than the chemically equilibrated HRG used in 
EOS~L and SM-EOS~Q (which has $e_\mathrm{dec}\eq0.143$\,GeV/fm$^3$ at the 
same temperature) \cite{Hirano:2002ds}, due to the larger-than-equilibrium
abundances of of baryon-antibaryon pairs and mesons that are prohibited
from annihilating as the system cools below $T_\mathrm{chem}$. So with 
s95p-PCE the fireball reaches the freeze-out point earlier, and it has a 
smaller freeze-out radius. It is this latter feature which causes the average 
radial flow along the freeze-out surface to be smaller for s95p-PCE than for 
the other two EoS: when plotting the radial velocity profiles along the 
decoupling surface, we found that all profiles are approximately linear 
functions of the radial distance $r$ from the center (qualitatively similar 
to the profiles shown in Fig.~4 of Ref.~\cite{Teaney:2001av}), and that the 
profile for s95p-PCE has the {\em largest} slope. However, the average radial 
flow is {\em smallest} because for s95p-PCE the average over the freeze-out
surface is truncated at a smaller maximal $r$ value.

%
\begin{figure*}[thb]
\includegraphics[width=\linewidth,clip=]{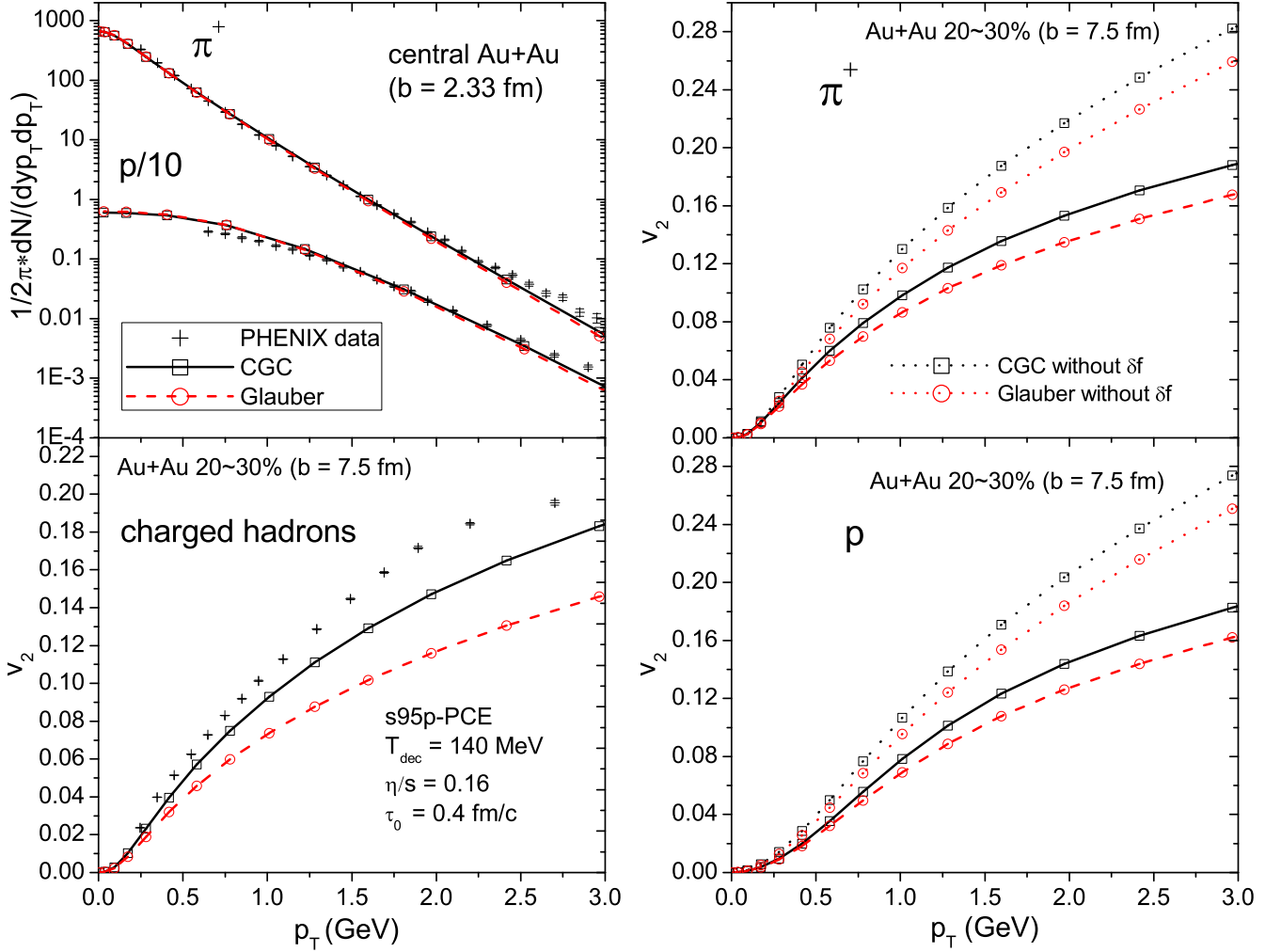}
\caption{Similar to Fig.~\ref{F4}, but for fixed $\eta/s\eq0.16$ and
different initial energy density profiles (Glauber vs. CGC) as 
indicated. The elliptic flow from an initial CGC density profile is
larger than for the Glauber initialization, due its larger initial 
eccentricity. 
}
\label{F8}
\end{figure*}
%

The charged hadron, pion and proton elliptic flows $v_2(p_T)$ show
quite large sensitivity to the EoS, especially at high $p_T$. (We repeat 
that the hydrodynamic spectra should probably not be trusted beyond
$p_T\sim 2-2.5$\,GeV/$c$, but plotting them out to 3\,GeV/$c$ makes
it easier to see what is going on in the calculation.) But we see that
most of this sensitivity comes in through the $\delta f$ correction at
freeze-out which is particularly large for SM-EOS~Q. The reason for this
is that the first-order phase transition leads to large velocity gradients
at the QGP-to-mixed-phase and mixed-phase-to-HRG interfaces 
\cite{Kolb:2000sd} which are largely but not completely washed out by 
viscous effects \cite{Song:2007ux} and leave traces on the decoupling 
surface. $\delta f$ effects are weaker with the smoother EOS~L than 
with SM-EOS~Q even though EOS~L generates on average more radial flow.

To discuss the contribution from collective flow anisotropies to pion
and proton $v_2(p_T)$ we focus on the dotted lines in the right panels
of Fig.~\ref{F7}. We see that, while s95p-PCE creates less radial
flow, it generates a larger flow {\em anisotropy} (we checked this
by direct computation), resulting in larger $v_2(p_T)$ for {\em both} 
pions and protons than with the other two equations of state. For EOS~Q 
it was found in \cite{Kolb:2002ve,Hirano:2002ds,Huovinen:2007xh} that 
if the kinetic freeze-out temperature $T_\mathrm{dec}$ is adjusted to
reproduce the $p_T$-spectra, the correct implementation of chemical 
freeze-out at $T_\mathrm{chem}$ in the HRG phase increases the 
mass-splitting between $v_2(p_T)$ of pions and protons at low $p_T$.  
On the other hand, if the freeze-out temperature is kept constant, the 
mass-splitting at low $p_T$ decreases \cite{Hirano:2002ds}. Since we 
have kept the freeze-out temperature fixed in our calculations, we see 
a similar phenomenon here: The elliptic flow mass splitting between 
pions and protons is weaker for the chemically frozen s95p-PCE than for 
the chemically equilibrated EOS~L and SM-EOS~Q. This is a consequence 
of the weaker radial flow generated by s95p-PCE.  

\subsection{Dependence on the shape of the initial energy density 
profile (CGC vs. Glauber)}
\label{sec4e}

We close with a discussion of the influence of the shape of the
initial energy density profile on the hadron spectra and elliptic 
flow, using the Glauber and CGC-fKLN models as examples. For an illustration
of these profiles see Fig.~\ref{F1}.

The CGC profile is characterized by slightly steeper normalized energy 
density gradients than the Glauber profile. According to the Euler 
equation for ideal fluids
\begin{equation}
  \dot{u}_\nu = \frac{c_s^2}{1+c_s^2}\frac{\nabla_{\!\nu} e}{e}
  \label{dudt}
\end{equation}
this leads to larger radial acceleration. Indeed, the upper left panel
of Fig.~\ref{F8} exhibits slightly flatter pion and proton spectra for
CGC-initialized simulations than for Glauber initial conditions.

The elliptic flow coefficients for charged hadrons, pions and protons
are all significantly larger for the CGC-initialized runs than for 
Glauber initial conditions. This is a direct consequence of the well-known
larger eccentricity of the CGC density profiles \cite{Hirano:2005xf,%
Drescher:2006pi,Luzum:2008cw,Heinz:2009cv} which drives a larger 
momentum anisotropy. The effect is qualitatively similar for all hadron 
species. The small amount of added radial flow from the CGC initialization 
that we see in the spectra has very little influence on the $p_T$-dependence
of the proton $v_2$ when compared to the much larger effects coming
from the larger source eccentricity. The suppression of $v_2$ by viscous
$\delta f$ corrections at freeze-out is similar for CGC and Glauber initial
conditions, being slightly larger in the CGC case. This is presumably
caused by the slightly larger flow velocities (and flow velicity 
gradients) generated by the CGC profile. 
   
\section{Conclusions}
\label{sec5}

We have performed a systematic study of the dependence of the pion and 
proton transverse momentum spectra and their $p_T$-dependent elliptic 
flow on the thermalization time $\tau_0$, initial energy density profile,
equation of state, freeze-out temperature and specific shear viscosity
in (2+1)-dimensional viscous hydrodynamic simulations. Assuming a temperature
independent shear viscosity to entropy ratio and CGC initial conditions
for the energy density profile, we find that the proton $p_T$-spectra 
measured in $200\,A$\,GeV central Au+Au collisions at RHIC 
favor $\eta/s$-values between 2 and 3 times the KSS bound 
$(\frac{\eta}{s})_\mathrm{KSS}\eq\frac{1}{4\pi}$ while the $p_T$-slope 
of the charged hadron elliptic flow prefers smaller values between 1 and
2 times the KSS bound. This tension cannot be resolved by different
choices for the other paramaters whose variation we studied. Of course,
the $\eta/s$ values extracted from a comparison with simulations using 
the less eccentric Glauber model for the initial energy density profile
are smaller, but the comparison with the experimental data gets worse
(the proton spectra come out steeper) and tension between the $\eta/s$
values preferred by spectra and $v_2$ gets stronger. Lower freeze-out
temperatures improve the agreement with the data, in particular with the
heavy-particle (proton) spectra. We saw very little sensitivity to the
choice of the termalization time $\tau_0$, but for larger values
of $\tau_0$ we did not allow for the evolution of pre-equilibrium 
radial and elliptic flow, contrary to what is expected to happen in 
reality. The main reason for not doing so was that, at this point, we 
have no theoretical control over this pre-equilibrium flow, and we did not
want to clutter our study by introducing still further parameters.
If there is a tendency worth mentioning in the context of varying
$\tau_0$ it is that smaller $\tau_0$-values lead to somewhat larger
radial flow which helps with the description of heavy hadron spectra.
This may, however, also be achievable by starting hydrodynamics later, 
but with non-zero initial transverse flow \cite{Broniowski:2008vp,%
Vredevoogd:2008id,Pratt:2008qv}. 

The main objective of this study was to gain an intuitive understanding
what reasonable changes in the key parameters of a viscous hydrodynamic
simulation will do to the final hadron spectra and elliptic flow. By
also keeping an eye on the available experimental data we come to the
conclusion that a purely hydrodynamic description of the experimental
spectra will probably not work, at least not with temperature independent
$\eta/s$. Realistic variations of $\eta/s$ with temperature are the subject
of a separate study \cite{ChunShen2010}. Based on that study combined with
the one presented here we believe that giving up on a (viscous) 
hydrodynamic description of the hadron resonance gas stage and replacing
it by a more reliable microscopic approach is unavoidable for a 
quantitative description of the experimental data. 

\appendix

\section{Analytic parametrization of EoS s95p-PCE \cite{s95p-PCE}}
\label{appa}

\begin{widetext}
We used the following analytic parametrization for the equation of state 
s95p-PCE (energy density $e$ and pressure $p$ in GeV/fm$^3$, entropy density 
$s$ in fm$^{-3}$, temperature $T$ in GeV):

\medskip
\noindent{\sl 1. Pressure:}
\begin{eqnarray}
p(e)\!\!&=&\!\!\left\{
\begin{array}{lcl}
 0.3299 \left[\exp(0.4346 e) - 1\right] & : & e < e_1\\ 
 1.024{\cdot}10^{{-}7}\cdot\exp(6.041 e) + 0.007273 + 0.14578 e 
                                  & : & e_1 < e < e_2\\
 0.30195 \exp(0.31308 e) - 0.256232  & : & e_2 < e < e_3\\
\!\! 0.332 e - 0.3223 e^{0.4585} - 0.003906 e{\cdot}\exp({-}0.05697 e) 
 + 0.1167 e^{{-}1.233} + 0.1436 e{\cdot}\exp({-}0.9131 e)            
                                  & : & e_3 < e < e_4\\
 0.3327 e - 0.3223 e^{0.4585} - 0.003906 e{\cdot}\exp({-}0.05697 e)
                                  & : & e > e_4
\end{array}
\right.\nonumber\\
\end{eqnarray}
where $e_1\eq0.5028563305441270$\,GeV/fm$^3$, 
      $e_2\eq1.62$\,GeV/fm$^3$,
      $e_3\eq1.86$\,GeV/fm$^3$, and
      $e_4\eq9.9878355786273545$ GeV/fm$^3$.

\medskip
\noindent{\sl 2. Entropy density:}
\begin{eqnarray}
s^{\frac{4}{3}}(e)\!\!&=&\!\!\left\{
\begin{array}{lcl}
 12.2304 e^{1.16849} & : &       e < e_1\\ 
 11.9279 e^{1.15635} & : & e_1 < e < e_2\\
 0.0580578 + 11.833 e^{1.16187} & : & e_2 < e < e_3\\\
 \!\!\!\!\!\left. \begin{array}{l}
 18.202 e - 62.021814 - 4.85479 \exp(-2.72407{\cdot}10^{-11} e^{4.54886})\\
 \ \ + 65.1272 e^{-0.128012} \exp(-0.00369624 e^{1.18735})
  - 4.75253 e^{-1.18423}
                \end{array} \right\}  & : & e_3 < e < e_4\\
 \!\!\!\left. \begin{array}{l}
 18.202 e - 63.0218 - 4.85479 \exp(-2.72407{\cdot}10^{-11} e^{4.54886})\\
 \ \  + 65.1272 e^{-0.128012} \exp(-0.00369624 e^{1.18735})
        \end{array} \right\} & : & e > e_4
                \end{array}\right.
\end{eqnarray}
where $e_1\eq0.1270769021427449$\,GeV/fm$^3$, 
      $e_2\eq0.4467079524674040$\,GeV/fm$^3$,
      $e_3\eq1.9402832534193788$\,GeV/fm$^3$, and
      $e_4\eq3.7292474570977285$\,GeV/fm$^3$.

\medskip
\noindent{\sl 3.Temperature:}
\begin{eqnarray}
T(e)\!\!&=&\!\!\left\{
\begin{array}{lcl}
 0.203054 e^{0.30679} & : & e < 0.5143939846236409\,\mathrm{GeV/fm}^3\\ 
 (e+p)/s              & : & e > 0.5143939846236409\,\mathrm{GeV/fm}^3
\end{array}
\right.
\end{eqnarray}

\end{widetext}


\acknowledgments{This work was supported by the U.S.\ Department of Energy 
under contracts DE-SC0004286 and DE-AC02-05CH11231 and within the framework 
of the JET Collaboration under grant number DE-SC0004104. P.H.'s research was
supported by the ExtreMe Matter Institute (EMMI). We thank Thomas Riley for 
helping us with the analytic parametrization of the EoS tables for s95p-PCE.}


\bibliographystyle{h-physrev3}

\bibliography{references}


\end{document}